\documentclass[conference]{IEEEtran}
\IEEEoverridecommandlockouts

\usepackage{cite}
\usepackage{amsmath,amssymb,amsfonts}
\usepackage{algorithmic}
\usepackage{graphicx}
\usepackage{textcomp}
\usepackage{xcolor}
\usepackage{booktabs}
\usepackage{url}
\usepackage{hyperref}

\usepackage[font=footnotesize]{caption}

\usepackage[belowskip=2pt,aboveskip=2pt]{caption}

\def\BibTeX{{\rm B\kern-.05em{\sc i\kern-.025em b}\kern-.08em
    T\kern-.1667em\lower.7ex\hbox{E}\kern-.125emX}}
\begin{document}

\title{
Towards Global Federated Genome-Wide Association Meta-Analysis Using GA4GH TES
}

\author{
\IEEEauthorblockN{
Abhijit Chunduru\IEEEauthorrefmark{1}\IEEEauthorrefmark{2}\IEEEauthorrefmark{3},
Matthew Joel\IEEEauthorrefmark{1}\IEEEauthorrefmark{3},
Zilinghan Li\IEEEauthorrefmark{1}\IEEEauthorrefmark{3},
Ravi Madduri\IEEEauthorrefmark{1}
}
\IEEEauthorblockA{
\IEEEauthorrefmark{1}Argonne National Laboratory 
\IEEEauthorrefmark{2}University of Massachusetts Amherst \\
\IEEEauthorrefmark{3}These authors contributed equally to this work.
}
\IEEEauthorblockA{
\{schunduru, mjoel, zilinghan.li, madduri\}@anl.gov
}
}


\maketitle

\begin{abstract}
Genome-wide association studies (GWAS) gain statistical power from large, ancestrally diverse cohorts, but privacy regulations and data-residency constraints often prevent genomic data from being centrally pooled across institutional or national borders. We present a privacy-preserving federated GWAS meta-analysis pipeline built on the APPFL framework, in which each site computes local GWAS summary statistics and transmits only aggregate results, never individual-level genotypes. Analysis is executed through a global network of Global Alliance for Genomics and Health (GA4GH) Task Execution Service (TES) endpoints, which allows computation to move to the data rather than the reverse. The server performs inverse-variance-weighted fixed-effect meta-analysis and returns aggregated results to all sites, while HiveWatch, our developed geographic observability toolkit, provides real-time monitoring of distributed task execution. In a five-site simulation over 100{,}000 synthetic individuals and roughly 240{,}000 variants for Type 2 Diabetes and Body Mass Index, the federated meta-analysis reproduces the association signal expected from a pooled analysis without centralizing any genotype data, showing that standards-based task execution and federated learning enables a practical privacy-preserving infrastructure for international GWAS meta-analysis.
\end{abstract}

\begin{IEEEkeywords}
Federated Learning, Privacy-Preserving Computation, GWAS Meta-Analysis, GA4GH Task Execution Service
\end{IEEEkeywords}

\vspace{-8pt}

\section{Introduction}
The statistical power of a genome-wide association study (GWAS) scales with cohort size and ancestral diversity, making large international consortia essential. However, the standard approach to centrally pooling individual-level genotype data is increasingly impractical due to privacy regulation, consent restrictions, and national data-residency rules that frequently prohibit export. The established solution is \emph{meta-analysis}: each site computes GWAS summary statistics locally and shares only per-variant aggregates, enabling a central server to produce joint results that are mathematically equivalent to a pooled fixed-effect analysis.
The remaining challenge lies in the \emph{infrastructure}: building a heterogeneous, multi-institutional, cross-border computing fabric around a common execution standard, where analysis executes within each data custodian's environment, raw genomic data never leaves their origin, and the distributed execution can be securely orchestrated and monitored.

In this paper, we present a pipeline that bridges the infrastructure gap by integrating three components: the Advanced Privacy-Preserving Federated Learning (APPFL) framework~\cite{ryu2022appfl,li2025advances} for the federation and aggregation; 
a network of Global Alliance for Genomics and Health (GA4GH) Task Execution Service (TES)~\cite{kanitz2024ga4gh} endpoints that implement the GA4GH standard for interoperable ``bring-compute-to-data'' task execution across heterogeneous computing infrastructure; and HiveWatch, a geographic observability toolkit developed by us for monitoring distributed task execution. Together, these components enable a federated GWAS meta-analysis over an international network of TES endpoints with real-time operational visibility.

\begin{figure}[!t]
    \centerline{\includegraphics[width=\columnwidth]{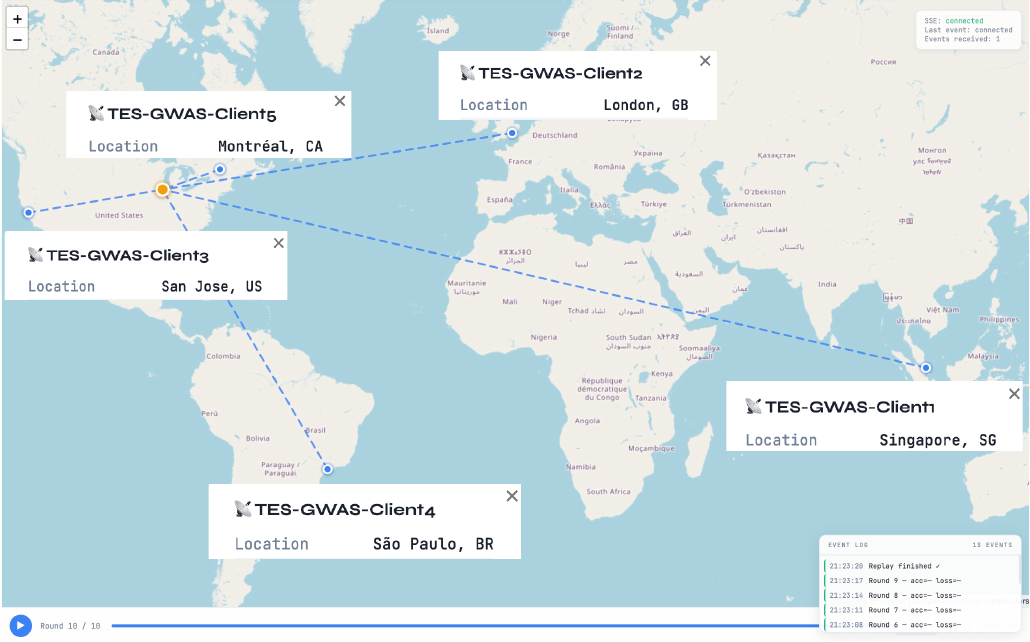}}
    \caption{International network of GA4GH TES endpoints. Each site runs a Funnel-based TES endpoint in a distinct cloud region; APPFL dispatches containerized GWAS tasks to the data, and only summary statistics are returned. HiveWatch monitors distributed computing in real-time.}
    \label{fig:map}
\end{figure}

\begin{figure*}[!htbp]
\centering
\includegraphics[width=0.95\linewidth]{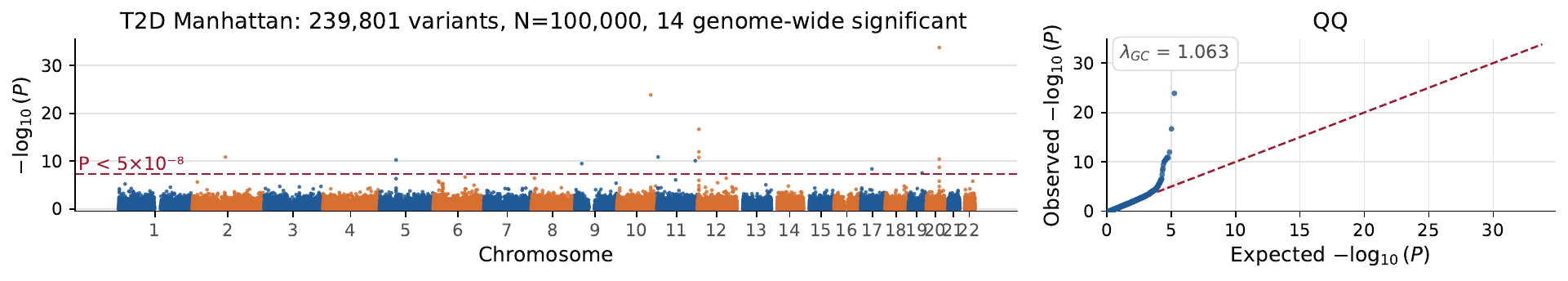}
\caption{Federated meta-analysis of type-2 diabetes across the five TES endpoints via
inverse-variance-weighted fixed-effect meta-analysis ($p<5\times10^{-8}$ threshold;
239{,}801 variants, $N=100{,}000$ partitioned as 18{,}032 / 9{,}237 / 25{,}028 / 40{,}752 /
6{,}951). Left: Manhattan plot, showing 14 genome-wide significant associations; the lead
signals at chr20:43.04~Mb and chr10:114.75~Mb correspond to the \textit{HNF4A} and
\textit{TCF7L2} loci. Right: quantile-quantile plot ($\lambda_{GC}=1.063$), in which observed
statistics follow the null expectation across the bulk of variants with departure confined to
the significant tail, indicating no inflation from the federated pipeline. Isolated rather
than clustered peaks reflect LD-pruning of the source genotypes.}
\vspace{-10pt}
\label{fig:results}
\end{figure*}

\vspace{-2pt}
\section{Pipeline Implementation}

\subsection{Federated Meta-Analysis}
The federated meta-analysis follows a standard cross-silo federated learning (FL) pattern in APPFL. Each site instantiates a \texttt{SiteGWASTrainer} that (i) loads local PLINK genotypes and phenotype/covariate tables, (ii) runs a chunked ordinary-least-squares GWAS for the quantitative trait (Body Mass Index) and a covariate-adjusted score test against a logistic null model for the binary trait (Type 2 Diabetes), and (iii) returns per-variant effect sizes $\beta_i$, standard errors $\mathrm{SE}_i$, minor-allele frequencies, and sample sizes. No individual-level genotype or phenotype ever leaves the site. The server runs a \texttt{MetaAnalysisAggregator} implementing inverse-variance-weighted fixed-effect meta-analysis. For each variant, with per-site weights $w_i = 1/\mathrm{SE}_i^2$,
\begin{equation}
\beta_{\mathrm{meta}} = \frac{\sum_i w_i\,\beta_i}{\sum_i w_i}, \qquad
\mathrm{SE}_{\mathrm{meta}} = \frac{1}{\sqrt{\sum_i w_i}},
\label{eq:ivw}
\end{equation}
from which a $Z$-statistic and $p$-value are derived. Because \eqref{eq:ivw} combines only transmitted aggregates, the federated estimate
asymptotically recovers the pooled fixed-effect result, coinciding exactly under
homogeneous per-site residual variance; genome-wide significance is assessed at
$p < 5\times10^{-8}$.

\subsection{TES Endpoint Network}
In our pipeline, each FL site launches a GA4GH TES endpoint, implemented with Funnel~\cite{funnel} and deployed on cloud compute in a distinct geographic region (Fig.~\ref{fig:map}). 
TES is a GA4GH standard that defines a common API for securely submitting, monitoring, and managing computational tasks across heterogeneous computing environments, enabling portable bring-compute-to-data workflows for genomics and other data-intensive biomedical applications~\cite{kanitz2024ga4gh}.
APPFL orchestrates federated GWAS by submitting containerized analysis tasks to each site's TES endpoint through the TES APIs for local execution. Each TES endpoint enforces authentication and authorization, allowing only the trusted federated server to submit, monitor, and retrieve the results of analysis tasks.
By exposing a standardized task execution interface, TES abstracts heterogeneous computing environments, allowing APPFL to interact with all participating sites through a common API while ensuring that individual-level genomic data remain local.

\subsection{Geographic Observability with HiveWatch}
To observe distributed execution across endpoints, we also developed HiveWatch (\href{https://github.com/APPFL/hivewatch}{https://github.com/APPFL/hivewatch}), a geographic observability toolkit integrated with the APPFL that records task dispatch, execution, and completion events and visualizes them on a live geographic map during the run (as shown in Fig.~\ref{fig:map}), with full replay from logs afterward. Endpoint-level visibility makes stalls, stragglers, and failures diagnosable across heterogeneous infrastructures.

\section{Evaluation and Results}
We evaluate the pipeline on a controlled simulation designed to have a known answer.
Synthetic European-ancestry genotypes for 100{,}000 individuals over 239{,}801 variants
(GRCh37), derived from 1000 Genomes reference structure, are partitioned across five sites of
deliberately unequal size (18{,}032; 9{,}237; 25{,}028; 40{,}752; and 6{,}951 individuals) to
exercise the sample-size weighting in \eqref{eq:ivw}. Phenotypes for Type 2 Diabetes and Body
Mass Index are simulated from PGS Catalog scores (PGS003443 and PGS004994) with age and sex
covariates. 

The federated GWAS meta-analysis recovers this structure without any genotype centralization
(Fig.~\ref{fig:results}). The Type 2 Diabetes scan yields 14 genome-wide significant
associations ($p<5\times10^{-8}$), led by chr20:43.04~Mb ($p=1.6\times10^{-34}$) and
chr10:114.75~Mb ($p=1.3\times10^{-24}$), which fall in \textit{HNF4A} and \textit{TCF7L2};
Body Mass Index yields 33 at the same threshold. The bulk of variants tracks the null diagonal
while true signal departs only in the tail ($\lambda_{GC}=1.063$ for T2D and $1.056$ for BMI),
indicating calibrated test statistics without inflation from the federation procedure. 

\section{Conclusion}
This work combines FL and GA4GH standards to enable deployable, privacy-preserving GWAS meta-analysis across international institutions. APPFL provides federated orchestration and exact fixed-effect meta-analysis, TES enables interoperable bring-compute-to-data execution across heterogeneous sites, and HiveWatch allows real-time observability. While our evaluation uses synthetic data to isolate the infrastructure contribution, future work will extend to real multi-institution deployments, more sophisticated meta-analysis models, stronger privacy guarantees, and resilient execution. 

\section*{Acknowledgment}
This material is based upon work supported by the U.S. Department of Energy, Office of Science, under contract number DE-AC02-06CH11357. We also gratefully acknowledge Amazon Web Services for providing cloud computing credits that were used to assist with experiments for this manuscript.

\bibliographystyle{IEEEtran}
\bibliography{references}
\end{document}